\documentclass{article}
       \renewenvironment{abstract}{\section*{Abstract}\small}{}
       \newtheorem{definition}{Definition}

       \newtheorem{theorem}[definition]{Theorem}
       
       \newtheorem{lemma}[definition]{Lemma}

       \newtheorem{proposition}[definition]{Proposition}

       \newtheorem{corollary}[definition]{Corollary}
       \makeatletter
       \renewcommand{\@begintheorem}[2]{ 
  \trivlist\item[\hskip\labelsep{\bf #1\ #2}]}
       \renewcommand{\@opargbegintheorem}[3]{\trivlist
         \item[\hskip \labelsep{\bf #1\ #2\ (#3)}]}
       \makeatother
       \newtheorem{proof}{Proof}

\usepackage{graphics}

\newcounter{fig}

\def\qed{\hfill\hbox{${\vcenter{\vbox{
        \hrule height 0.4pt\hbox{\vrule width 0.4pt height 6pt
        \kern5pt\vrule width 0.4pt}\hrule height 0.4pt}}}$}}

\title{Program schemes with binary write-once arrays and the complexity classes they capture}

\author{Iain A. Stewart\thanks{Supported by EPSRC Grants GR/K 96564 and GR/M
12933.},\\ Department of Mathematics and Computer Science,\\
University of Leicester, Leicester LE1 7RH, U.K.}

\date{}

\begin{document}

\maketitle

\begin{abstract}  We study a class of program schemes, NPSB, in which,
aside from basic assignments, non-deterministic guessing and
while loops, we have access to arrays; but where these arrays are
binary write-once in that they are initialized to `zero' and can
only ever be set to `one'. We show, amongst other results, that:
NPSB can be realized as a vectorized Lindstr\"{o}m logic; there
are problems accepted by program schemes of NPSB that are not
definable in the bounded-variable infinitary logic ${\cal
L}^\omega_{\infty\omega}$; all problems accepted by the program
schemes of NPSB have a zero-one law; and on ordered structures,
NPSB captures the complexity class $\mbox{{\bf
L}}^{\mbox{\scriptsize{\bf NP}\normalsize}}$. The class of
program schemes NPSB is actually the union of an infinite
hierarchy of classes of program schemes. When we amend the
semantics of our program schemes slightly, we find that the
classes of the resulting hierarchy capture the complexity classes
$\Sigma^p_i$ (where $i\geq 1$) of the Polynomial Hierarchy {\bf
PH}. Finally, we give logical equivalences of the
complexity-theoretic question `Does {\bf NP} equal {\bf PSPACE}?'
where the logics (and classes of program schemes) involved define
only problems with zero-one laws (and so do not define some
computationally trivial problems).\end{abstract}

\section{Introduction}

Finite model theory is essentially the study of logical
definability over finite structures. An important sub-area of
finite model theory is the relationship between the logical
definability of classes of finite structures and computational
complexity theory; that is, descriptive complexity theory. This
relationship is best exemplified by Fagin's seminal result that a
problem, i.e., a class of finite structures over the same
signature, can be defined by a sentence of existential
second-order logic if, and only if, the problem (or, to be more
precise, an encoding of it) can be accepted by a polynomial-time
non-deterministic Turing machine \cite{Fag74}.

In two recent papers \cite{ACS99,SteCSL}, we have studied logical
definability in a slightly different context in that we have
worked with classes of program schemes as opposed to more
traditional logics. Program schemes are more computational in
flavour than are formulae of traditional logics yet they remain
amenable to logical manipulation. The concept of a program scheme
originates from the 1970's with work of, for example, Constable,
Friedman, Gries, Hewitt and Paterson \cite{CG72,Fri71,PH70}; and
complexity-theoretic considerations of such program schemes were
subsequently studied by, for example, Harel, Jones, Muchnik,
Peleg, Tiuryn and Urzyczyn \cite{HP84,JM77,TU88}. Our analysis of
program schemes differs from what has gone before in that we are
always concerned with finite structures (and not infinite ones as
was often the case previously) and we do not assume that the
elements of our finite structures are necessarily linearly
ordered. Our studies have exhibited an extremely close
relationship between certain classes of program schemes and the
more traditional logics from descriptive complexity theory, and
our methodology has enabled us to prove new results extending
known results from descriptive complexity theory. For example, in
\cite{ACS99} we defined a hierarchy of classes of program schemes,
NPSS, where these program schemes involve assignments, while
loops and non-deterministic guessing and have access to a stack,
before showing that this hierarchy is proper and (the union of
it) has exactly the same expressive power as path system logic (a
logic previously considered on ordered finite structures
\cite{LSS98,Ste94} and which has been shown to be none other than
stratified fixed point logic and stratified Datalog on the class
of all finite structures \cite{Gro97,Kol91}). Furthermore, our
hierarchy result translates to a strong hierarchy result about
path system logic, a result which was previously unknown and not
immediately derivable using known results and techniques from
descriptive complexity theory (see \cite{ACS99} for more details).

As well as equating classes of program schemes with more
mainstream logics from descriptive complexity theory, we have
also shown how naturally defined classes of program schemes
result in logics which previously have not arisen in descriptive
complexity theory yet which have interesting properties. For
example, in \cite{SteCSL} we considered a hierarchy of classes of
program schemes, NPSA, similar to the program schemes of NPSS but
with arrays replacing the stack (with the levels of the hierarchy
denoted by $\mbox{NPSA}(i)$, for $i\geq 1$). We showed that (the
union of) this hierarchy NPSA can be realized as a vectorized
Lindstr\"{o}m logic (many of the mainstream logics of descriptive
complexity theory are explicitly defined as vectorized
Lindstr\"{o}m logics: see \cite{EF95}), and that there are program
schemes in this hierarchy (even in the first level) which accept
problems not definable in the bounded-variable infinitary logic
${\cal L}^\omega_{\infty\omega}$ (a fundamental and encompassing
logic in finite model theory: see \cite{EF95}). However, we also
show that, like any problem definable in ${\cal
L}^\omega_{\infty\omega}$, every problem accepted by a program
scheme of NPSA has a zero-one law. We remark that on ordered
finite structures the class of program schemes NPSS captures the
complexity class {\bf P} and the class of program schemes NPSA
captures the complexity class {\bf PSPACE} (and in both cases the
underlying hierarchy collapses to the first level).

In this paper, in an attempt to `bridge the gap' between the
polynomial-time world of NPSS and the polynomial-space world of
NPSA, we modify the program schemes of NPSA so that all arrays are
`binary write-once' in the sense that all array elements are
initially set at `zero' and the only modification to any array
element allowed is to set it to `one'. The resulting class of
program schemes is denoted NPSB, with the levels of the underlying
hierarchy being $\mbox{NPSB}(i)$, for $i\geq 1$. We show that
NPSB retains some of the properties of NPSA: like NPSA, NPSB can
be realized as a vectorized Lindstr\"{o}m logic, and
$\mbox{NPSB}(1)$ contains problems not definable in ${\cal
L}^\omega_{\infty\omega}$ (every problem accepted by a program
scheme of NPSB trivially has a zero-one law as NPSB is a
sub-class of NPSA); but whereas both NPSA and $\mbox{NPSA}(1)$
capture {\bf PSPACE} on ordered structures, $\mbox{NPSB}(1)$
captures {\bf NP} and NPSB captures $\mbox{{\bf
L}}^{\mbox{\scriptsize{\bf NP}\normalsize}}$, with the NPSB
hierarchy collapsing to $\mbox{NPSB}(3)$.

We then amend the semantics of the class of program schemes NPSB
in that we allow the current values of arrays to be `passed
across' to other program schemes (appearing as, what amounts to,
subroutines in the main program scheme) in a computation
(hitherto, the semantics has only allowed the current values of
variables to be passed across). We denote the class of program
schemes with this amended semantics as $\mbox{NPSB}^p$ (the
subscript reflects the polynomially many values passed across to
the component program schemes). We show that on the class of all
finite structures, $\mbox{NPSB}^p(2i-1)=\mbox{NPSB}^p(2i)$ and
captures $\Sigma_i^p$, i.e., the $i$th level of the Polynomial
Hierarchy {\bf PH}, for $i\geq 2$; and so $\mbox{NPSB}^p$
captures {\bf PH} itself.

Finally, we compare the relative expressibilities of the classes
of program schemes $\mbox{NPSA}(1)$ and $\mbox{NPSB}(1)$. Recall
that any problem accepted by any program scheme of these classes
has a zero-one law; and so there are complexity-theoretic trivial
problems (like the problem consisting of all those structures of
even size over some fixed signature) not accepted by such program
schemes. We show that $\mbox{NPSA}(1) = \mbox{{\bf PSPACE}}\cap
{\cal EXT}$ and $\mbox{NPSB}(1) = \mbox{{\bf NP}}\cap {\cal EXT}$,
where ${\cal EXT}$ is the class of problems closed under
extensions (and thus every non-trivial problem in ${\cal EXT}$ has
a `one-law'); and thus that the question of whether {\bf NP} is
equal to {\bf PSPACE} is equivalent to the question of whether
the two classes of program schemes $\mbox{NPSB}(1)$ and
$\mbox{NPSA}(1)$ accept the same class of problems (and also
equivalent to whether two particular fragments of two vectorized
Lindstr\"{o}m logics are equally expressive) on the class of all
finite structures.

This paper is structured as follows. In the next section, we
outline the definitions relating to this paper before we define
our classes of program schemes in Section 3. In Section 4, we
identify the class of program schemes NPSB with a vectorized
Lindstr\"{o}m logic, before looking at our program schemes and
logics on ordered structures in Section 5, where we also amend
the original semantics of our program schemes. In Section 6, we
compare the relative computational power of the classes of
program schemes $\mbox{NPSB}(1)$ and $\mbox{NPSA}(1)$ (on the
class of all finite structures) before presenting our conclusions
and directions for further research in Section 7.

\section{Preliminaries}

The main reference texts for the basic concepts, notions and
results of finite model theory are \cite{EF95,Imm98} and it is to
these books that we refer the reader for additional information.
However, we give the definitions relating to this paper in some
detail below as it is often the case that more generality is
required than in \cite{EF95,Imm98} and we also need some notions
not defined in those texts.

Ordinarily, a {\em signature\/}
$\sigma$ is a tuple
$\langle R_1,\ldots,R_r,C_1,\ldots,C_c\rangle$, where each $R_i$ is a relation
symbol, of arity $a_i$, and each $C_j$ is a constant symbol. {\em First-order
logic over the signature $\sigma$\/}, $\mbox{FO}(\sigma)$, consists of those
formulae built from atomic formulae over $\sigma$ using $\wedge$, $\vee$, $\neg$,
$\forall$ and
$\exists$; and $\mbox{FO}=\cup\{\mbox{FO}(\sigma): \sigma \mbox{ is some
signature}\}$.

A {\em finite structure ${\cal A}$ over the signature $\sigma$\/}, or {\em
$\sigma$-structure\/}, consists of a finite {\em universe\/} or {\em domain\/}
$|{\cal A}|$ together with a relation $R_i$ of arity $a_i$, for every relation
symbol $R_i$ of $\sigma$, and a constant $C_j\in|{\cal A}|$, for every constant
symbol $C_j$ (by an abuse of notation, we do not usually distinguish between
constants and relations, $C_j^{{\cal A}}$ and $R_i^{{\cal A}}$, and constant and
relation symbols, $C_j$ and $R_i$). A finite structure
${\cal A}$ whose domain consists of $n$ distinct elements has {\em size\/} $n$,
and we denote the size of ${\cal A}$ by $|{\cal A}|$ also (this does not cause
confusion). We only ever consider finite structures of size at least 2, and the
set of all finite structures of size at least 2 over the signature $\sigma$ is
denoted $\mbox{STRUCT}(\sigma)$. A {\em problem} over some signature
$\sigma$ consists of a subset of $\mbox{STRUCT}(\sigma)$ that is closed under
isomorphism; that is, if ${\cal A}$ is in the problem then so is every isomorphic
copy of ${\cal A}$. Throughout, all our structures are finite.

The class of problems defined by the sentences of FO is denoted by
FO also, and we do likewise for other logics. It is widely
acknowledged that, as a means for defining problems, first-order
logic leaves a lot to be desired; especially when we have in mind
developing a relationship between computational complexity and
logical definability. For example, every first-order definable
problem can be accepted by a log-space deterministic Turing
machine (where structures are encoded as strings) yet there are
problems in the complexity class {\bf L} (log-space) which can not
be defined in first-order logic (one such being the problem
consisting of all those structures, over any signature, that have
even size). Consequently, a number of methods have been developed
so as to increase definability.

One method is to extend first-order logic using a
{\em vectorized sequence of Lindstr\"om quantifiers\/} corresponding to some
problem $\Omega$; or, as we prefer, an {\em operator\/} $\Omega$ for short. Suppose
that $\Omega$ is over the signature $\sigma$, where
$\sigma=\langle R_1, \ldots, R_r, C_1, \ldots, C_c\rangle$, as above. The logic
$(\pm\Omega)^\ast[\mbox{FO}]$ consists of those formulae built using the usual
constructs of first-order logic and also the operator $\Omega$, where the operator
$\Omega$ is applied as follows.
\begin{itemize}
\item Suppose that $\psi_1({\bf x}_1,{\bf y}),\ldots,\psi_r({\bf x}_r,{\bf y})$
are formulae of $(\pm\Omega)^\ast[\mbox{FO}]$ such that:
\begin{itemize}
\item each ${\bf x}_i$ is a $ka_i$-tuple of distinct variables, for some
fixed $k\geq 1$;
\item ${\bf y}$ is an $m$-tuple of distinct variables, for some $m\geq 0$, each of
which is different from any variable of ${\bf x}_1,\ldots,{\bf x}_r$; and
\item all free variables of any $\psi_i$ are contained in either ${\bf x}_i$ or
${\bf y}$.
\end{itemize}
\item Suppose that ${\bf d}_1,\ldots,{\bf d}_c$ are $k$-tuples of variables and
constants (which need not be distinct).
\item Then:
$$\Omega[\lambda {\bf x}_1\psi({\bf x}_1,{\bf y}),\ldots,{\bf x}_r\psi_r({\bf
x}_r,{\bf y})]({\bf d}_1,\ldots,{\bf d}_c)$$
is a formula of $(\pm\Omega)^\ast[\mbox{FO}]$ whose free variables are the
variables of ${\bf y}$ together with any other variables appearing in ${\bf
d}_1,\ldots,{\bf d}_c$.
\end{itemize}
If $\Phi$ is a sentence of the form $\Omega[\lambda {\bf
x}_1\psi_1({\bf x}_1),\ldots,{\bf x}_r\psi_r({\bf x}_r)]({\bf
d}_1,\ldots,{\bf d}_c)$, as above, over some signature
$\sigma^\prime$ then we interpret $\Phi$ in a
$\sigma^\prime$-structure ${\cal A}$ as follows (note that as
$\Phi$ is a sentence, the variables of ${\bf y}$ are absent and
the tuples ${\bf d}_1,\ldots,{\bf d}_c$, which are only there if
there are constant symbols in $\sigma$, consist entirely of
constant symbols of $\sigma^\prime$).
\begin{itemize}
\item The domain of the $\sigma$-structure $\Phi({\cal A})$ is
$|{\cal A}|^k$.
\item The relation $R_i$ of $\Phi({\cal A})$ is defined via:
\begin{itemize}
\item for any ${\bf
u}\in|\Phi({\cal A})|^{a_i}=|{\cal A}|^{ka_i}$, $R_i({\bf u})$ holds in $\Phi({\cal
A})$ if, and only if,
$\psi_i({\bf u})$ holds in ${\cal A}$.
\end{itemize}
\item The constant $C_j$ of $\Phi({\cal A})$ is defined via:
\begin{itemize}
\item $C_j$ is the
interpretation of the tuple of constants ${\bf d}_j$ in ${\cal A}$.
\end{itemize}
\end{itemize}
We define that ${\cal A}\models\Phi$ if, and only if, $\Phi({\cal A})\in\Omega$
(the situation where $\Phi$ has free variables is similar except that $\Phi$ is
interpreted in expansions of $\sigma^\prime$-structures by an appropriate number of
constants). We call logics such as $(\pm\Omega)^\ast[\mbox{FO}]$ {\em
vectorized Lindstr\"om logics\/}. We shall also be interested in fragments of
vectorized Lindstr\"om logics where: the formulae are such that the operator
$\Omega$ does not appear within the scope of a negation sign,
namely $\Omega^\ast[\mbox{FO}]$ (the {\em positive\/} fragment of
$(\pm\Omega)^\ast[\mbox{FO}]$); and further the formulae are such that there are
no nestings of the operator $\Omega$, namely
$\Omega^1[\mbox{FO}]$ (the {\em positive unnested\/} fragment of
$(\pm\Omega)^\ast[\mbox{FO}]$).

It can be the case that (a fragment of) a vectorized Lindstr\"om
logic $(\pm\Omega)^\ast[\mbox{FO}]$ has a very straightforward
normal form; a normal form which obviates the need to nest
applications of the operator $\Omega$ and which tells us something
about the `degree of difficulty' of the particular problem
$\Omega$ with respect to the class of problems defined by the
sentences of the logic. For example, suppose that every problem in
(a fragment of) $(\pm\Omega)^\ast[\mbox{FO}]$ can be defined by a
sentence of the form $\Omega[\lambda {\bf x}_1\psi_1({\bf
x}_1),\ldots,{\bf x}_r\psi_r({\bf x}_r)]({\bf
d}_1,$\linebreak$\ldots,{\bf d}_c)$, as above, except where each
$\psi_i$ is quantifier-free first-order. Then we say that the
problem $\Omega$ is {\em complete\/} for (the fragment of)
$(\pm\Omega)^\ast[\mbox{FO}]$ via {\em quantifier-free first-order
translations\/}. This is directly analogous to completeness for
some complexity class via some resource-bounded reduction: in
fact, as we shall see, such normal forms can often yield very
strong complexity-theoretic completeness results.

Vectorized Lindstr\"om logics have been studied quite extensively
in finite model theory and a whole range of complexity classes
have been {\em captured\/}, i.e., characterized, by vectorized
Lindstr\"om logics (see, for example,
\cite{Got97,Imm87,SteJLC1,SteHP} and the references therein).
However, some (though not all) of these characterizations only
hold in the presence of a built-in successor relation. Consider
some vectorized Lindstr\"om logic $(\pm\Omega)^\ast[\mbox{FO}]$.
To say that this logic has a {\em built-in successor relation\/},
which we denote by $(\pm\Omega)^\ast[\mbox{FO}_s]$, means that no
matter which signature $\sigma^\prime$ we are working over, there
is always a binary relation symbol $succ$ and two constant symbols
$0$ and $max$ available (none of which is in $\sigma^\prime$) such
that $succ$ is always interpreted as a successor relation with
least element $0$ and greatest element $max$ in any
$\sigma^\prime$-structure. That is, for any
$\sigma^\prime$-structure of size $n$, $succ$ is always of the
form $\{(0,u_1),(u_1,u_2),\ldots,(u_{n-2},max)\}$, where the
elements of $\{0,u_1,u_2,\ldots,u_{n-2},max\}$ are distinct.
However, there is a further semantic stipulation on the sentences
of $(\pm\Omega)^\ast[\mbox{FO}_s]$: we only consider as
well-formed those sentences for which the interpretation in any
structure is independent of the particular successor relation
chosen. For example, define the problem TC over the signature
$\sigma_{2++}=\langle E,C,D\rangle$, where $E$ is a binary
relation symbol and $C$ and $D$ are constant symbols, as
consisting of all those $\sigma_{2++}$-structures for which, when
considered as digraphs in the natural way, there is a directed
path from the vertex $C$ to the vertex $D$. Then the following
sentence is a well-formed sentence of
$(\pm\mbox{TC})^\ast[\mbox{FO}_s]$ (as satisfiability in a given
structure is invariant with respect to $succ$):
\begin{eqnarray*}
\lefteqn{\mbox{TC}[\lambda (x_1,x_2),(y_1,y_2)(x_1=0\wedge y_1=max\wedge
succ(x_2,y_2))}\\ & &
\hspace{1in}\vee (x_1=max\wedge y_1=0\wedge succ(x_2,y_2))](0,0,max,max),
\end{eqnarray*}
and it defines the problem over the empty signature consisting of
those structures of even size. (Note that in \cite{EF95}, the
mechanism by which a successor relation is introduced into a
logic is slightly different from how we have described here in
that only problems on {\em ordered structures\/} are ever
considered: see \cite{EF95}. Nevertheless, the two approaches
essentially amount to the same thing and we shall refer to the
two mechanisms interchangeably). Note also that other relations
can be built into logics in the same way as is a successor
relation; or even just two distinct constants can be built in.)

From a logical perspective, there is a problem with our built-in
successor relation in the following sense. Given a sentence of
first-order logic in which the relation symbol $succ$ appears
(and in which other constant and relation symbols might appear),
it is actually undecidable as to whether the sentence is
invariant with respect to $succ$. That is, there does not exist
an effective enumeration of the well-formed sentences of
$\mbox{FO}_s$. Given this fact, it is highly debatable as to
whether any `logic' $(\pm\Omega)^\ast[\mbox{FO}_s]$ should really
be called a logic; and it is an open question currently occupying
much research activity as to whether there actually exists a
logic capturing the complexity class {\bf P} (polynomial-time),
or indeed any complexity class contained in {\bf NP}
(non-deterministic polynomial-time), where by `contained in' we
really mean `contained in but expected to be different from' (as
{\bf NP} itself can be captured by a logic, one such being
existential second-order logic). The reader is referred to
\cite{Ott97} for more details on this and related points.
(Notwithstanding the above discussion, we still refer to
$(\pm\Omega)^\ast[\mbox{FO}_s]$ as a logic on the grounds of
convenience.)

Theorem~\ref{CUB=NP}, below, is an example of a normal form
result. Define the problem CUB, over the signature
$\sigma_2=\langle E\rangle$, where $E$ is a binary relation
symbol, as follows.
\begin{eqnarray*}
\mbox{CUB} = \{{\cal G}\in\mbox{STRUCT}(\sigma_2) & : & \mbox{the graph } {\cal G}
\mbox{ has a subset of edges inducing a}\\ & & \mbox{regular subgraph of degree
3}\}
\end{eqnarray*} (think of a $\sigma_2$-structure ${\cal G}$ as encoding an
undirected graph via: `there is an edge $(u,v)$ if, and only if,
$u \neq v$ and $E(u,v)\vee E(v,u)$ holds in ${\cal G}$'). We shall
need the following result later on.

\begin{theorem}\label{CUB=NP}\cite{SteTCS}
{\em The complexity class {\bf NP} is identical to the class of
problems defined by the sentences of
$\mbox{CUB\/}^\ast[\mbox{FO\/}_s]$\/{\em ;} and any problem in
{\bf NP} can be defined by a sentence of
$\mbox{CUB\/}^1[\mbox{FO\/}_s]$ of the form\/{\em :}
$$\mbox{CUB\/}[\lambda{\bf x},{\bf y}\psi({\bf x},{\bf y})],$$
where $|{\bf x}|=|{\bf y}|=k$, for some $k\geq 1$, and $\psi$ is
a quantifier-free formula of $\mbox{FO\/}_s$. Hence,
$\mbox{CUB\/}^\ast[\mbox{FO\/}_s]=\mbox{CUB\/}^1[\mbox{FO\/}_s]=\mbox{{\bf
NP}}$ and CUB is complete for {\bf NP} via quantifier-free
first-order translations with successor.}\qed
\end{theorem}

Note that Theorem~\ref{CUB=NP} subsumes the `traditional' known
complexity-theoretic result that CUB is complete for {\bf NP} via
log-space reductions (a result attributed to Chv\`atal in
\cite{GJ79}).

\section{Program schemes}

Program schemes are more `computational' means for defining classes of problems
than are logical formulae. A {\em program scheme\/} $\rho\in\mbox{NPSA}(1)$
involves a finite set
$\{x_1, x_2, \ldots,x_k\}$ of {\em variables\/}, for some $k\geq 1$, and is over a
signature $\sigma$. It consists of a finite sequence of {\em instructions\/} where
each instruction, apart from the first and the last, is one of the following:
\begin{itemize}
\item[$\bullet$] an {\em assignment instruction\/} of the form `{\tt $x_i$ :=
$y$}', where
$i\in\{1,2,\ldots,k\}$ and where
$y$ is a variable from $\{x_1,x_2,\ldots,x_k\}$, a constant symbol of
$\sigma$ or one of the special constant symbols $0$ and $max$ which do not appear
in any signature;
\item[$\bullet$] an {\em assignment instruction\/} of the form `{\tt $x_i$ :=
$A[y_1,y_2,\ldots,y_d]$}' or `{\tt $A[y_1,y_2,\ldots,$}\linebreak{\tt $y_d]$ :=
$y_0$}', for some
$i\in\{1,2,\ldots,k\}$, where each
$y_j$ is a variable from $\{x_1,x_2,$\linebreak$\ldots,x_k\}$, a constant symbol of
$\sigma$ or one of the special constant symbols $0$ and $max$ which do not appear
in any signature, and where $A$ is an array symbol of dimension $d$;
\item[$\bullet$] a {\em guess instruction\/} of the form `{\tt guess $x_i$}', where
$i\in\{1,2,\ldots,k\}$; or
\item[$\bullet$] a {\em while instruction\/} of the form `{\tt while $\varphi$ do
$\alpha_1;\alpha_2;\ldots; \alpha_q$ od}', where $\varphi$ is a
quantifier-free formula of $\mbox{FO}(\sigma\cup\{0,max\})$ whose
free variables are from $\{x_1,x_2,\ldots,x_k\}$, and where each
of $\alpha_1, \alpha_2, \ldots, \alpha_q$ is another instruction
of one of the forms given here (note that there may be nested
while instructions).
\end{itemize} The first instruction of $\rho$ is `{\tt
input$(x_1,x_2,\ldots,x_l)$}' and the last instruction is `{\tt
output}\linebreak$(x_1, x_2, \ldots, x_l)$', for some
$l$ where
$1\leq l\leq k$. The variables $x_1,x_2,\ldots,x_l$ are the {\em input-output
variables\/} of
$\rho$, the variables $x_{l+1}, x_{l+2}, \ldots,x_k$ are the {\em free
variables\/} of
$\rho$ and, further, any free variable of $\rho$ never appears on the left-hand
side of an assignment instruction nor in a guess instruction. Essentially, free
variables appear in $\rho$ as if they were constant symbols.

A program scheme $\rho\in\mbox{NPSA}(1)$ over $\sigma$ with $t$ free variables,
say, takes a
$\sigma$-structure ${\cal A}$ and $t$ additional values from $|{\cal A}|$, one for
each free variable of
$\rho$, as input; that is, an expansion ${\cal A}^\prime$ of ${\cal A}$ by
adjoining
$t$ additional constants. The program scheme $\rho$ computes on
${\cal A}^\prime$ in the obvious way except that:
\begin{itemize}
\item[$\bullet$] execution of the instruction `{\tt guess $x_i$}'
non-deterministically assigns an element of $|{\cal A}|$ to the variable $x_i$;
\item[$\bullet$] the constants $0$ and $max$ are interpreted as two arbitrary but
distinct elements of
$|{\cal A}|$; and
\item[$\bullet$] initially, every input-output variable and every array element is
assumed to have the value $0$.
\end{itemize} Note that throughout a computation of $\rho$, the value of any free
variable does not change.  The expansion
${\cal A}^\prime$ of the structure
${\cal A}$ is {\em accepted\/} by
$\rho$, and we write ${\cal A}^\prime\models\rho$, if, and only if, there exists a
computation of $\rho$ on this expansion such that the output-instruction is reached
with all input-output variables having the value $max$ (in particular, some
computations might not be terminating). We can easily build the usual `if' and
`if-then-else' instructions using while instructions (see, for example,
\cite{SteActa}). Henceforth, we shall assume that these instructions are at our
disposal.

We want the sets of structures accepted by our program schemes to
be problems, i.e., closed under isomorphism, and so we only ever
consider program schemes $\rho$ where a structure is accepted by
$\rho$ when $0$ and $max$ are given two distinct values from the
universe of the structure if, and only if, it is accepted no
matter which pair of distinct values is chosen for $0$ and $max$.
This is analogous to how we build two constant symbols into a
logic. Furthermore, we can build a successor relation into the
program schemes of $\mbox{NPSA}(1)$ so as to obtain the class of
program schemes $\mbox{NPSA}_s(1)$. As with our logics, we write
$\mbox{NPSA}(1)$ and $\mbox{NPSA}_s(1)$ to also denote the class
of problems accepted by the program schemes of $\mbox{NPSA}(1)$
and $\mbox{NPSA}_s(1)$, respectively (and do likewise with other
classes of program schemes).

We have two remarks. First, our notation $\mbox{NPSA}(1)$ reflects
the fact that $\mbox{NPSA}(1)$ is the first level of an infinite
hierarchy of classes of program schemes, as we shall see
presently. Second, as the definition of our class of program
schemes $\mbox{NPSA}(1)$ stands, we do not know whether the
program schemes in this class can be recursively enumerated.
However, we are prepared to live with this (possible)
inconvenience as we could have defined the program schemes of
$\mbox{NPSA}(1)$ to be devoid of the constant symbols $0$ and
$max$ and be such that initially every variable and array element
is non-deterministically assigned the same element of the input
structure (this would result in the same class of problems). Such
a definition would mean that the class of structures accepted by
such a program scheme is always closed under isomorphism (hence,
recursive enumerability would not be an issue). However, there are
three real reasons for having the constant symbols $0$ and $max$
in our program schemes. First, we can use $0$ to initialize all
variables and arrays, with the result that we never have to worry
about whether an assignment involves an uninitialized variable or
array element. Second, having two distinct constants around is
useful when it comes to programming. Third, we shall soon use the
constant symbol $max$ to enable us to study `binary write-once'
arrays (that is, arrays where the elements can only be set to
$max$ and thereafter remain unchanged).

Henceforth, we think of our program schemes as being written in the style of a
computer program. That is, each instruction is written on one line and while
instructions (and, similarly, if and if-then-else instructions) are split so that
`{\tt while $\varphi$ do}' appears on one line, `$\alpha_1$' appears on the next,
`$\alpha_2$' on the next, and so on (of course, if any $\alpha_i$ is a while, if or
if-then-else instruction then it is split over a number of lines in the same way).
The instructions are labelled 1, 2, and so on, according to the line they appear
on. In particular, every instruction is considered to be an assignment, a guess or
a test. An {\em instantaneous description \/{\em (}ID\/{\em )}} of a program
scheme on some input consists of a value for each variable,  the number of the
instruction about to be executed and values for all array elements. A {\em partial
ID\/} consists of just a value for each variable and the number of the instruction
about to be executed. One {\em step\/} in a program scheme computation is the
execution of one instruction, which takes one ID to another, and we say that a
program scheme can {\em move\/} from one ID to another if there exists a sequence
of steps taking the former ID to the latter.

As we hinted at above, the class of program schemes
$\mbox{NPSA}(1)$ is but the first level of an infinite hierarchy
of program schemes. Suppose that we have defined a class of
program schemes $\mbox{NPSA}(2m-1)$, for some $m\geq 1$, and that
any program scheme has associated with it: a set of input-output
variables; a set of free variables; and a set of bound variables
(this is certainly the case when $m=1$, where the associated set
of bound variables is empty).

\begin{definition}\label{NPSA2m}
Let the program scheme $\rho\in\mbox{NPSA}(2m-1)$ be over the
signature $\sigma$. Suppose that $\rho$ has: input-output
variables $x_1, x_2, \ldots, x_k$; free variables $x_{k+1},
x_{k+2},$\linebreak$\ldots, x_{k+s}$; and bound variables
$x_{k+s+1}, x_{k+s+2},\ldots, x_{k+s+t}$. Let
$x_{i_1},x_{i_2},\ldots,x_{i_p}$ be free variables of $\rho$, for
some $p$ (and so $k+1 \leq i_1< i_2 < \ldots < i_p \leq k+s$).
Then: $$\forall x_{i_1}\forall x_{i_2}\ldots\forall x_{i_p}\rho$$
is a program scheme of $\mbox{NPSA}(2m)$, which we denote by
$\rho^\prime$, with: no input-output variables; free variables
those of
$\{x_{k+1},x_{k+2},\ldots,x_{k+s}\}\setminus\{x_{i_1},x_{i_2},\ldots,x_{i_p}\}$;
and the remaining variables of $\{x_1,x_2,\ldots,x_{k+s+t}\}$ as
its bound variables.

A program scheme such as $\rho^\prime$ takes expansions
$\mathcal{A}^\prime$ of $\sigma$-structures $\mathcal{A}$ by
adjoining $s-p$ constants as input (one for each free variable),
and $\rho^\prime$ accepts such an expansion $\mathcal{A}^\prime$
if, and only if, for every expansion $\mathcal{A}^{\prime\prime}$
of $\mathcal{A}^\prime$ by $p$ additional constants (one for each
variable $x_{i_j}$, for $j\in\{1,2,\ldots,p\}$),
$\mathcal{A}^{\prime\prime}\models\rho$ (the computation on such
an expansion $\mathcal{A}^{\prime\prime}$ always starts with the
arrays initialised to $0$).\qed
\end{definition}

\begin{definition}\label{NPSA2m-1} A program scheme $\rho^\prime
\in\mbox{NPSA}(2m-1)$, for some $m\geq 2$, over the signature
$\sigma$, is defined exactly as is a program scheme of NPSA$(1)$
except that the test in any while instruction is a program scheme
$\rho\in\mbox{NPSA}(2m-2)$. The bound variables of $\rho^\prime$
consist of the bound variables of any test in any while
instruction; all free variables in any test in any while
instruction are input-output or free variables of $\rho^\prime$;
and there may be other free and input-output variables (appearing
in $\rho^\prime$ at the `top level' but not in any test). Of
course, any free variable never appears on the left-hand side of
an assignment instruction or in a guess instruction (at the `top
level').

Suppose that a program scheme $\rho^\prime\in\mbox{NPSA}(2m-1)$
has $s$ free variables. Then it takes an expansion
$\mathcal{A}^\prime$ of a $\sigma$-structure $\mathcal{A}$ by
adjoining $s$ constants as input and computes on
$\mathcal{A}^\prime$ in the obvious way; except that when some
while instruction is encountered, the test, which is a program
scheme $\rho\in\mbox{NPSA}(2m-2)$, is evaluated according to the
expansion of $\mathcal{A}^\prime$ by the current values of any
relevant input-output variables of $\rho^\prime$ (which may be
free in $\rho$). In order to evaluate this test, the arrays
associated with $\rho$ are initialized at $0$ and when the test
has been evaluated the computation of $\rho^\prime$ resumes
accordingly with the values of its arrays and input-output and
free variables being exactly as they were immediately prior to the
test being evaluated. In particular, array values can not be
`passed across' in the evaluation of tests: the values of
variables can be but they are never amended in the process. \qed
\end{definition}

Consequently, we obtain a hierarchy of classes of problems:
$$\mbox{NPSA}(1) \subseteq \mbox{NPSA}(2) \subseteq \ldots
\subseteq \cup\{\mbox{NPSA}(i): i=1,2, \ldots\}=\mbox{NPSA}$$ (we
use the inclusion relation between consecutive classes because
this is how they are related as classes of problems). It is easy
to see that, for one thing, $\mbox{FO}\subseteq\mbox{NPSA}$.

In this paper, we are primarily interested in some sub-classes of
program schemes of $\mbox{NPSA}$, namely the sub-classes
$\mbox{NPSB}(i)$, for $i=1,2,\ldots$, and the union of these
classes NPSB, where the only allowed assignment instructions with
an array element on the left-hand side are of the form {\tt
$A[x_1,x_2,\ldots,x_k]$ := $max$}; that is, the only values array
elements can have are $0$ and $max$, and once an array element is
set to $max$ then it remains at $max$ thereafter (the notation
reflects the binary nature of these arrays). Obviously,
$\mbox{NPSB}(i)\subseteq\mbox{NPSA}(i)$, for all $i=1,2,\ldots$;
and $\mbox{NPSB}\subseteq\mbox{NPSA}$.

Results concerning the program schemes of NPSA have already been
obtained, and some of these results relevant to this paper are
stated below. A problem $\Omega$, over some signature $\sigma$
and where the domain of any $\sigma$-structure of size $n$ is
taken to be $\{1,2,\ldots,n\}$, for which the function $f(n)$,
defined as the number of structures in $\Omega$ of size $n$
divided by the number of $\sigma$-structures of size $n$, is such
that the limit as $n$ tends to infinity exists and is 0 or 1 is
said to have a {\em zero-one law\/}.

\begin{theorem}\label{NPSAmain}\cite{SteCSL}
\begin{itemize}
\item[({\em i\/})] {\em There exists a problem $\Omega_a$, involving reachability
in Petri nets, for which $$\mbox{NPSA} =
(\pm\Omega_a)^\ast[\mbox{FO}],$$ and the class of problems NPSA
has a zero-one law.}
\item[({\em ii\/})] {\em There is a quantifier-free
first-order translation with \/{\em 2} constants from any problem
in $\mbox{NPSA\/}(1)$ to the problem $\Omega_a$\/{\em ;} and so
$\Omega_a$ is complete for $\mbox{NPSA\/}(1)$ via quantifier-free
first-order translations with \/{\em 2} constants.}
\item[({\em iii\/})] {\em The problem CUB is in NPSA$(1)$ but not definable in the logic $\cal{L}^\omega_{\infty\omega}$.}
\item[({\em iv\/})] {\em In the presence of a built-in successor
relation, the hierarchy $\mbox{NPSA}_s$ collapses to the first
level, $\mbox{NPSA}_s(1)$, and captures the complexity class {\bf
PSPACE}.}\qed
\end{itemize}\end{theorem}

It is worth mentioning the role of the logic ${\cal
L}^\omega_{\infty\omega}$ in finite model theory. This logic is
an important logic for a number of reasons, one of which is that
it subsumes many of the logics from finite model theory
(including transitive-closure logic, least fixed point logic and
partial fixed point logic) in that these logics can be realized
as fragments of ${\cal L}^\omega_{\infty\omega}$. Furthermore,
${\cal L}^\omega_{\infty\omega}$ has a zero-one law and so any
logic subsumed by ${\cal L}^\omega_{\infty\omega}$ has a zero-one
law. It is particularly interesting that $\mbox{NPSA}(1)$ (and so
also NPSA) can not be realized as a fragment of ${\cal
L}^\omega_{\infty\omega}$ (as CUB is a problem in
$\mbox{NPSA}(1)$ that is not in ${\cal L}^\omega_{\infty\omega}$:
a result proven in \cite{SteMLQ}).

In the absence of arrays, when the resulting class of program
schemes is denoted NPS, and additionally in the presence of a
stack, when the resulting class of program schemes is denoted
NPSS, there are results analogous to parts ({\em i\/}), ({\em
ii\/}) and ({\em iv\/}) of Theorem~\ref{NPSAmain} (see
\cite{ACS99}) in that: both NPS and NPSS can be realized as
vectorized Lindstr\"{o}m logics so that the problems
corresponding to the operators involved in these logics are
complete for NPS$(1)$ and NPSS$(1)$ via quantifier-free
first-order translations with 2 constants; and on ordered
structures, the complexity classes captured are {\bf NL}
(non-deterministic log-space) and {\bf P}, respectively. However,
unlike NPSA, both NPS and NPSS can be realized as fragments of
${\cal L}^\omega_{\infty\omega}$. Furthermore, the underlying
hierarchies of NPS and NPSS are proper at every level (even if we
restrict to problems only involving trees) whereas, as we shall
affirm later, all that is known as regards NPSA is that
$\mbox{NPSA}(1) \subset \mbox{NPSA}(2) \subset \mbox{NPSA}(3)$.

\section{Partitioned Petri nets}

We begin by describing a generalization of the digraph
reachability problem to a scenario where the moves between nodes
depend upon the availability and utilization of external
resources. We first describe the basic decision problem in an
everyday fashion before we consider a manifestation of it as a
class of structures over a given signature and see how this
problem is related to computation in the program schemes of
$\mbox{NPSB}(1)$.

Consider the following scenario. We are given a direct graph
$G=(V,E)$, where $|V| = n$, with a source vertex $source$ and a
sink vertex $sink$, but where each edge is labelled with a
(possibly empty) set of labels with each label being of one of
the following forms:
\begin{itemize}
\item `user resource $r_i$ is unused';
\item `system resource $s_j$ is available';
\item `user resource $r_i$ is unused and this move uses this
resource but makes the system resource $s_j$ available (if it
wasn't available previously)'.
\end{itemize}
There is a polynomial number of different {\em user resources\/}
$\{r_i : i = 1,2,\ldots,p(n)\}$, which are either in the state
`used' or the state `unused'; and a polynomial number of {\em
system resources\/} $\{s_i : i = 1,2,\ldots,q(n)\}$, which are
either in the state `available' or the state `unavailable' (for
some polynomials $p$ and $q$). A move in the digraph from vertex
$u$ to vertex $v$ via the edge $(u,v)$ can only be made if either
no labels label the edge $(u,v)$ or at least one of the labels
labelling the edge $(u,v)$ is satisfied (with a resulting change
in the state of a user resource, and possibly a system resource,
if the label is of the third type). The question we ask is, given
the initial state where all user resources are unused and no
system resources are available, is it possible to move from
$source$ to $sink$ in our given environment? That is, can the
user use his or her resources wisely so as to enable a traversal
in the digraph from the source to the sink?

Note that whether a move can be made depends only on certain
predicates involving the states of the resources: for example,
there are no moves dependent upon the state of a user resource
being `used' or of a system resource being `unavailable'. The
situation is as it is as this decision problem arises naturally
from our consideration of our program schemes; but we comment
further on this problem and related problems in the Conclusion.

We encode the above decision problem as a problem, i.e., class of
finite structures, involving Petri nets. Our encoding is natural
and has certain properties which we shall utilize later. The
reader is referred to \cite{EN94} for the basic notions and
concepts relating to Petri nets (this reference also gives details
of numerous complexity-theoretic results concerning fundamental
problems in Petri nets).

\begin{definition}\label{partPetri} Define $\sigma_b=\langle
P,Q,T_1,T_2,T_3,C,D\rangle$ where $P$, $Q$, $T_1$, $T_2$ and
$T_3$ are relation symbols of arities 1, 1, 2, 3 and 4,
respectively, and $C$ and $D$ are constant symbols. Let ${\cal
P}$ be a $\sigma_b$-structure. We can think of the elements of
$|{\cal P}|$ as being the places of a Petri net and the relations
$P$ and $Q$ as describing two partitions of these places. We can
think of:
\begin{itemize}
\item the relation $T_1$ as describing the set of transitions $$\{(\{u\},\{v\}):
u,v\in P \mbox{ and } T_1(u,v) \mbox{ holds}\};$$
\item the relation $T_2$ as describing the set of transitions
\begin{eqnarray*}
\lefteqn{\{(\{u,i\},\{v,i\}): u,v\in P, i\not\in P, i\in Q \mbox{
and } T_2(u,v,i) \mbox{ holds}\}}\\ & & \cup \{(\{u,j\},\{v,j\}):
u,v\in P, j\not\in P,Q \mbox{ and } T_2(u,v,j) \mbox{
holds}\};\hspace{0.5in}\end{eqnarray*}and
\item the relation $T_3$ as describing the set of transitions
$$\{(\{u,i\},\{v,j\}): u,v\in P, i,j\not\in P,i\in Q,
j\not\in Q \mbox{ and } T_3(u,v,i,j) \mbox{ holds}\}.$$
\end{itemize} Furthermore, the initial marking of our Petri
consists of the place $C$ and the places not in $P$ but in $Q$. We
define the problem $\Omega_b$ as
\begin{eqnarray*}
\{{\cal P}\in\mbox{STRUCT}(\sigma_b) & : & \mbox{there is a marking reachable from
the initial marking}\\ & & \mbox{in which there is at least one token on the place
} D\}.\hspace{2mm}\qed
\end{eqnarray*}
\end{definition}

Note that the transitions encoded within a $\sigma_b$-structure
${\cal P}$ are of one of four types, as depicted in
\refstepcounter{fig}\label{transitions}Fig.~\thefig, and that the
relations $T_1$, $T_2$ and $T_3$ of ${\cal P}$ might have
additional tuples in them that do not affect how we think of
${\cal P}$ as a Petri net.

\begin{center}
\includegraphics{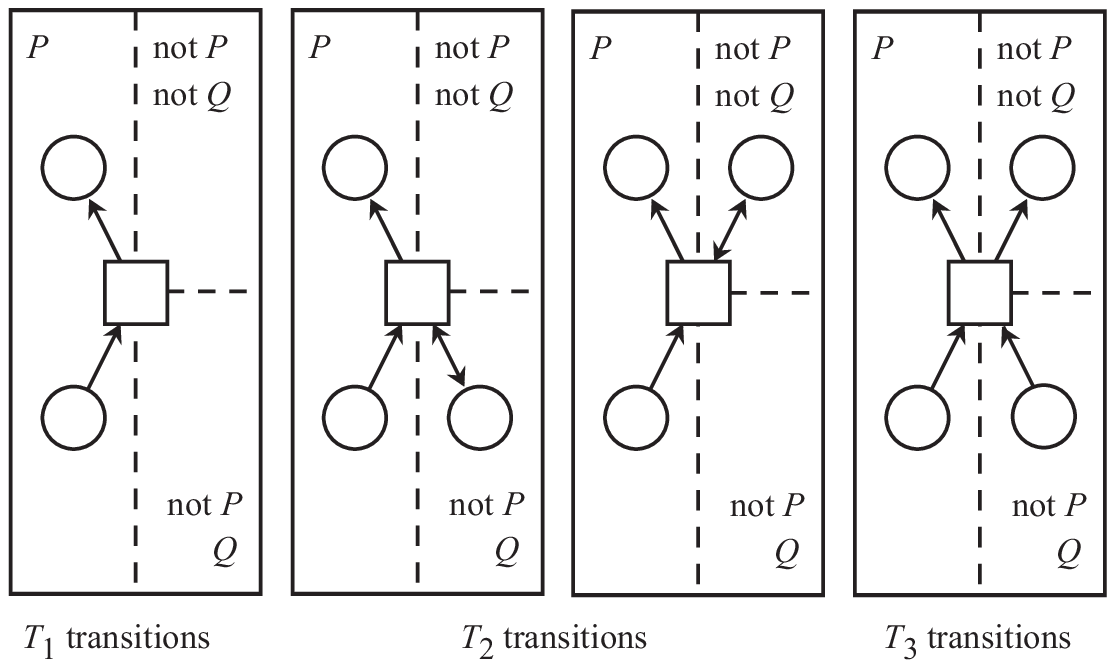}

Figure~\ref{transitions}. The different types of transitions.
\end{center}

With reference to our decision problem presented earlier, it
should be clear that: the places in $P$ correspond to the
vertices $V$ of our digraph $G=(V,E)$, with $C$ corresponding to
the source vertex and $D$ the sink vertex; the places not in $P$
but in $Q$ correspond to the user resources; and the places not
in $P$ and not in $Q$ correspond to the system resources
(henceforth, we shall use this terminology to describe the places
of our Petri net). Additionally, the transitions described by
$T_1$ correspond to edges of $E$ with no labels; the transitions
described by $T_2$ yield edges labelled with labels of the form
`user resource $r_i$ is unused' and `system resource $s_j$ is
available'; and the transitions described by $T_3$ yield edges
labelled with labels of the form `user resource $r_i$ is unused
and this move uses this resource but makes the system resource
$s_j$ available'. We interpret a user resource as being in the
state `unused', if there is a token on it, and as being in the
state `used' otherwise (such places only ever have at most one
token on them). It may be the case, in a reachable marking, that
a system resource has more than one token on it. However, tokens
can not be removed from such places. Thus, it is only ever
important as to whether a system resource has no tokens on it,
when we think of it being in the state `unavailable', or at least
one token on it when, we think of it being in the state
`available'.

The proof of the following theorem is similar to those in
\cite{SteCSL} although there are additional complications caused
by only having assignments which set array values to $max$.

\begin{theorem}\label{NPSB(1)=Omega_b}{\em There is a quantifier-free first-order
translation with \/{\em 2} constants from any problem in $\mbox{NPSB\/}(1)$ to the
problem $\Omega_b$. Hence, $\Omega_b$ is complete for $\mbox{NPSB\/}(1)$ via
quantifier-free first-order translations with
\/{\em 2} constants. }\end{theorem}

\begin{proof} Let $\rho$ be a program scheme of $\mbox{NPSB}(1)$ over some
signature $\sigma$ in which if and if-then-else instructions
might occur. W.l.o.g., we may assume that array symbols only
appear in assignment instructions, that there is only one array
symbol, $B$, and that this array symbol has dimension $d\geq 1$.
We assume that the variables involved in $\rho$ are
$x_1,x_2,\ldots,x_k$.

Let ${\cal A}$ be a $\sigma$-structure of size $n\geq 2$. An element
${\bf u}=(u_0,u_1,\ldots,u_k)$ of $\{1,2,\ldots,l\}\times|{\cal A}|^k$ encodes a
partial ID of $\rho$ on input ${\cal A}$ via: a computation of $\rho$ on
${\cal A}$ is about to execute instruction $u_0$ and the variables
$x_1,x_2,\ldots,x_k$ currently have the values $u_1,u_2,\ldots,u_k$, respectively.
Henceforth, we identify partial IDs of $\rho$ and the elements of
$\{1,2,\ldots,l\}\times|{\cal A}|^k$.

We now build a Petri net ${\cal P}$, as in
Definition~\ref{partPetri}, using $\rho$ and ${\cal A}$. Our
Petri net ${\cal P}$ has a set of places consisting of the set
$\{1,2,\ldots,l\}\times|{\cal A}|^k$ in union with the set
$\{{\bf w}_0,{\bf w}_m:{\bf w}\in|{\cal A}|^d\}$. The sets of
places $P$ and $Q$ are
$$P=\{1,2,\ldots,l\}\times|{\cal A}|^k \mbox{ and } Q=\{{\bf w}_0:{\bf w}\in|{\cal
A}|^d\},$$ respectively. Hence, the user resources are $\{{\bf
w}_0:{\bf w}\in|{\cal A}|^d\}$ and the system resources $\{{\bf
w}_m:{\bf w}\in|{\cal A}|^d\}$. We shall use a token on the user
resource $(w_1,w_2,\dots,w_d)_0$ to signify that the current
value of $B[w_1,w_2,\dots,w_d]$ is $0$; and a token on the system
resource $(w_1,w_2,\dots,w_d)_m$ to signify that the current
value of $B[w_1,w_2,\dots,w_d]$ is $max$. Obviously, we have to
take care to ensure that a marking does not yield contradictory
interpretations.

Let ${\bf u}\in\{1,2,\ldots,l\}\times|{\cal A}|^k$.

Suppose that the instruction $u_0$ does not involve the array
symbol $B$ and it is possible for $\rho$ on input ${\cal A}$ to
move from any ID whose partial ID is ${\bf u}$ to an ID whose
partial ID is ${\bf v}$ in one step. Then the transition $(\{{\bf
u}\},\{{\bf v}\})$ is in $T_1$ (more precisely, the pair $({\bf
u},{\bf v})$ is in $T_1$).

Suppose that the instruction $u_0$ is of the form
$x_j:=B[x_{i_1},x_{i_2},\ldots,x_{i_d}]$ and it is possible for
$\rho$ on input ${\cal A}$ to move from any ID whose partial ID
is ${\bf u}$ to an ID whose partial ID is ${\bf v}$ in one step
(because the value of $B[x_{i_1},x_{i_2},\ldots,x_{i_d}]$ is such
that $\rho$ on input ${\cal A}$ can move from an ID whose partial
ID is ${\bf u}$ to an ID whose partial ID is ${\bf v}$ in one
step). Then both of the transitions $(\{{\bf
u},(u_{i_1},u_{i_2},\ldots,u_{i_d})_0\},\{{\bf
v},(u_{i_1},u_{i_2},\ldots,u_{i_d})_0\})$ and $(\{{\bf
u},(u_{i_1},u_{i_2},\ldots,u_{i_d})_m\},\{{\bf
v},(u_{i_1},u_{i_2},\ldots,u_{i_d})_m\})$ are in $T_2$ (of
course, in the former transition, $v_j$ is 0, and in the latter
$v_j$ is $max$, with $u_i=v_i$, for all $i=1,2,\ldots,k$ different
from $j$).

Suppose that the instruction $u_0$ is of the form
$B[x_{i_1},x_{i_2},\ldots,x_{i_d}]:= max$ and it is possible for
$\rho$ on input ${\cal A}$ to move from an ID whose partial ID is
${\bf u}$ to an ID whose partial ID is ${\bf v}$ in one step.
Then the transition $(\{{\bf
u},(u_{i_1},u_{i_2},\ldots,u_{i_d})_m\},\{{\bf
v},(u_{i_1},u_{i_2},\ldots,$\linebreak$u_{i_d})_m\})$ is in $T_2$
and the transition $(\{{\bf
u},(u_{i_1},u_{i_2},\ldots,u_{i_d})_0\},\{{\bf
v},(u_{i_1},u_{i_2},\ldots,u_{i_d})_m\})$ is in $T_3$ (of course,
in both transitions $u_i = v_i$, for $i=1,2,\ldots,k$).

Our initial marking of ${\cal P}$ is such that there is one token
on each place of $\{{\bf w}_0: {\bf w}\in|{\cal A}|^d\}$ and one
token on the place $(1,{\bf 0})\in\{1,2,\ldots,l\}\times|{\cal
A}|^k$, which we define to be $C$; and we define $D$ as the place
$(l,{\bf max})\in\{1,2,\ldots,l\}\times|{\cal A}|^k$.

It is not difficult to see that our Petri net ${\cal P}$ (that
is, our $\sigma_b$-structure ${\cal P}$) can be described in
terms of the $\sigma$-structure ${\cal A}$ using quantifier-free
first-order formulae (in which $0$ and $max$ appear: explicit
descriptions of structures by quantifier-free first-order
formulae are given in, for example, \cite{SteTCS0}).
Consequently, in order for the result to follow we need to show
that: ${\cal A}\models\rho$ if, and only if, ${\cal
P}\in\Omega_b$; and that $\Omega_b\in\mbox{NPSB}(1)$.

Suppose that ${\cal A}\models\rho$. Then there is a sequence $\pi$
of (full, not partial) IDs starting at the initial ID (where all
variables have the value 0, where the instruction to be executed
is instruction 1 and where the array $B$ has the value 0
throughout) and ending in a final ID (where all variables have the
value $max$ and where the instruction to be executed is
instruction $l$) such that $\rho$ moves from one ID in $\pi$ to
the next in one step. As hinted earlier, we can mirror any ID with
a set of markings of our Petri net ${\cal P}$ as follows. If the
ID consists of the partial ID ${\bf u}\in\{1,2,\ldots,l\}\times
|{\cal A}|^k$ together with some valuation on the array $B$ then
the place ${\bf u}$ is marked with one token as are the places of
$\{{\bf w}_0:{\bf w}\in|{\cal A}|^d,B[{\bf w}]=0\}$, and the
places of $\{{\bf w}_m:{\bf w}\in|{\cal A}|^d,B[{\bf w}]=max\}$
are marked with at least one token. This accounts for all tokens.
Note that the initial ID of $\rho$ corresponds to the initial
marking of ${\cal P}$. A simple analysis yields that if $\rho$ on
input ${\cal A}$ moves from one ID to another in one step then the
Petri net can fire a transition to move from the marking
corresponding to the first ID to a marking corresponding to the
subsequent ID; and conversely (as remarked earlier, as regards
the system resources, it does not matter how many tokens reside on
them but only whether or not at least one token resides). Hence,
${\cal A}\models\rho$ if, and only if, ${\cal P}\in\Omega_b$.

All that remains is to show that $\Omega_b\in\mbox{NPSB}(1)$.
There are two essential difficulties in deriving a program scheme
to accept $\Omega_b$. First, a $\sigma_b$-structure ${\cal P}$
might be such that a reachable marking involves more than one
token on some system resource; and we need to cater for this
event when we simulate a sequence of transitions in ${\cal P}$ by
an execution of a program scheme on input ${\cal P}$. Second, we
need to keep track of where tokens are in a way which avoids us
modelling the fact that a token is on a place simply by using an
array indexed by the place names; for we are not allowed to
register that a token has moved from a place by assigning some
array element the value $0$ (recall, the only assignment
instruction allowed on an array element is to set that element to
$max$).

Our Petri net ${\cal P}$ is such that initially there is one
token, call it $t$, on the place $C$ of $P$ and there is one
token on every user resource (we assume that the place $C$ is
indeed in $P$: otherwise, our program scheme simply rejects the
input ${\cal P}$). No other tokens are involved in the initial
marking. Also, transitions are such that we can imagine the token
$t$ as being moved from place to place amongst the places of $P$,
and we can imagine every other token either staying where it is,
after some transition, or being moved from user resource to a
system resource, and then staying where it is thereafter.

As regards our first difficulty, we do not need to actually
monitor how many tokens lie on any system resource but only
whether there is at least one token such a place. This obviates
the need to count tokens. As regards our second difficulty, in
order to decide whether (at least) one token lies on some system
resource $s$, we use a dedicated array $B_1$, of dimension 1, so
that whenever a token is placed on such a $s$ then $B_1[s]$ is set
at $max$: once $B_1[s]$ has been set to $max$ we know that there
will be a token on $s$ thereafter. In order to decide whether a
token lies on some user resource $r$, we use an array $B_2$, of
dimension 1, to register when the token originally on the place
$r$ is first moved from $r$ by setting $B_2[r]$ equal to $max$ at
this point. Consequently, if we wish to know whether there is a
token on such a place $r$, we test to see whether $B_2[r]=0$
holds. Finally, we model the movement of the solitary token $t$ by
using a dedicated variable, $x$ say: that is, the token $t$ is on
place $p$ if, and only if, $x$ has the value $p$. Given the above
discussion, it is straightforward to see that the problem
$\Omega_b$ can be accepted by a program scheme of
$\mbox{NPSB}(1)$, and so the result follows.\qed
\end{proof}

In essence, Theorem~\ref{NPSB(1)=Omega_b} tells us that any
problem accepted by a program scheme of $\mbox{NPSB}(1)$ can be
described by a sentence of the form
\begin{eqnarray*}
\lefteqn{\Omega_b[\lambda {\bf x}\psi_P({\bf x}),{\bf
x}\psi_Q({\bf x}),{\bf x},{\bf y}\psi_1({\bf x},{\bf y}), {\bf
x},{\bf y},{\bf z}\psi_2({\bf x},{\bf y},{\bf z}),}\\ & &
\hspace{2.5in} {\bf x},{\bf y},{\bf z},{\bf w} \psi_3({\bf
x},{\bf y},{\bf z},{\bf w})]({\bf u},{\bf v}),
\end{eqnarray*} where: $|{\bf x}|=|{\bf y}|=|{\bf z}|=|{\bf w}|=k$, for some $k\geq 1$, and all variables
are distinct; $\psi_P$, $\psi_Q$, $\psi_1$, $\psi_2$ and $\psi_3$
are quantifier-free first-order formulae over
$\sigma_b\cup\{0,max\}$; and ${\bf u}$ and ${\bf v}$ are
$k$-tuples of constant symbols (in fact, we can actually take
${\bf u}$ to be $0$ repeated $k$ times and ${\bf v}$ to be $max$
repeated $k$ times: moreover, the sentence is such that whether it
is true in some given structure is independent of the distinct
values chosen for $0$ and $max$).

Similarly to as in \cite{SteCSL}, Theorem~\ref{NPSB(1)=Omega_b}
allows us to relate the class of problems accepted by the program
schemes of NPSB with the class of problems defined by the
sentences of the logic $(\pm\Omega_b)^\ast[\mbox{FO}]$. For each
$m\geq 1$, we define the fragment $\pm\Omega_b(m)$ of
$(\pm\Omega_b)^\ast[\mbox{FO}]$ as follows.
\begin{itemize}
\item $\pm\Omega_b(1)$ consists of all formulae of the form
$$\Omega_b[\lambda {\bf x}\psi_P,{\bf
x}\psi_Q,{\bf x},{\bf y}\psi_1, {\bf x},{\bf y},{\bf
z}\psi_2,{\bf x},{\bf y},{\bf z},{\bf w} \psi_3]({\bf u},{\bf
v}),$$ where: $\psi_P$, $\psi_Q$, $\psi_1$, $\psi_2$ and $\psi_3$
are quantifier-free first-order formulae over
$\sigma_b\cup\{0,max\}$; ${\bf u}$ and ${\bf v}$ are $k$-tuples
of constant symbols or variables; there may be other free
variables; and the truth of any interpretation of the formula
(over a relevant structure and with values given for any free
variables) is independent of the pair of distinct values chosen
for $0$ and $max$.
\item
$\pm\Omega_b(m+1)$, for odd $m\geq 1$, consists of the universal
closure of $\pm\Omega_b(m)$; that is, the set of formulae of the
form $\forall z_1\forall z_2\ldots\forall z_k \psi,$ where $\psi$
is a formula of $\pm\Omega_b(m)$.
\item $\pm\Omega_b(m+1)$, for even $m\geq 2$, consists of the set of formulae of
the form
\begin{eqnarray*}\lefteqn{\Omega_b[\lambda {\bf x}(\psi^1_P\vee \neg \psi^2_P),{\bf
x}(\psi^1_Q\vee \neg\psi^2_Q),{\bf x},{\bf
y}(\psi^1_1\vee\neg\psi_1^2), {\bf x},{\bf y},{\bf
z}(\psi^1_2\vee\neg\psi_2^2),}\\ & & \hspace{2.5in} {\bf x},{\bf
y},{\bf z},{\bf w} (\psi^1_3\vee\neg\psi_3^2)]({\bf u},{\bf
v}),\end{eqnarray*}where: $\psi^1_P$, $\psi^2_P$, $\psi^1_Q$,
$\psi^2_Q$, $\psi^1_1$, $\psi^2_1$, $\psi^1_2$, $\psi^2_2$,
$\psi^1_3$ and $\psi^2_3$ are formulae of $\pm\Omega_b(m)$; ${\bf
u}$ and ${\bf v}$ are tuples of constant symbols or variables;
there may be other free variables; and the truth of any
interpretation of the formula (over a relevant structure and with
values given for any free variables) is independent of the pair
of distinct values chosen for $0$ and $max$.
\end{itemize}
As in \cite{SteTCS}, a straightforward induction yields that:
\begin{itemize}
\item for every odd $m\geq 1$, every formula in the closure of $\pm\Omega_b(m)$
under $\wedge$, $\vee$ and $\exists$ is logically equivalent to a
formula of $\pm\Omega_b(m)$; and
\item for every even $m\geq 1$, every formula in the closure of $\pm\Omega_b(m)$
under $\wedge$, $\vee$ and $\forall$ is logically equivalent to a
formula of $\pm\Omega_b(m)$.
\end{itemize}
Consequently, $(\pm\Omega_b)^\ast[\mbox{FO}]=\cup\{\Omega_b(m):
m\geq 1\}$.

\begin{corollary}\label{Omegab=NPSB} {\em In the presence of $2$ built-in constant
symbols, $\pm\Omega_b(m)=\mbox{NPSB\/}(m)$, for each $m\geq
1$\/{\em ;} and so $(\pm
\Omega_b)^\ast[\mbox{FO\/}]=\mbox{NPSB\/}$. }
\end{corollary}

\begin{proof} We proceed by induction on $m$ almost identically to the proof of Corollary 10 from \cite{SteCSL}.
The base case, when $m=1$, follows by
Theorem~\ref{NPSB(1)=Omega_b}.\qed
\end{proof}

Note that $(\pm \Omega_b)^\ast[\mbox{FO}]=\mbox{NPSB}$ even in
the absence of our 2 built-in constant symbols as we can `build
them ourselves' using existential quantification.

We end this section by showing that NPSB can not be realized as a
fragment of ${\cal L}^\omega_{\infty\omega}$ (unlike NPS and
NPSS).

\begin{lemma}\label{CUBinNPSB} {\em The problem CUB can be accepted by a program
scheme of $\mbox{NPSB\/}(1)$.}
\end{lemma}

\begin{proof} It was shown in \cite{SteCSL} that CUB is in
$\mbox{NPSA}(1)$: however, the program scheme used there to
accept CUB is not in $\mbox{NPSB}(1)$. Nevertheless, the basic
approach can be amended to yield a program scheme of
$\mbox{NPSB}(1)$.

Let ${\cal G}$ be a $\sigma_2$-structure. We begin by `guessing'
a set of distinct edges in the graph ${\cal G}$. We use two
3-dimensional array symbols, $B_1$ and $B_2$, to store these
guessed edges. In particular, if our first guessed edge is
$(u_1,v_1)$, having checked that $(u_1,v_1)$ is indeed an edge of
${\cal G}$, we set $B_1[0,0,u_1]=max$ and $B_2[0,0,v_1]=max$.
Next, we guess an edge $(u_2,v_2)$, check to see whether this
edge is indeed an edge of ${\cal G}$ and then check to see
whether this edge is different from $(u_1,v_1)$. If so then we
set $B_1[u_1,v_1,u_2]=max$ and $B_2[u_1,v_1,v_2]=max$: otherwise,
we set $B_1[u_1,v_1,max]=max$ and $B_2[u_1,v_1,max]=max$ and stop
guessing. We continue in this fashion until the guessing stage
stops whence we have a list of distinct edges of ${\cal G}$.

Finally, we check to see whether the guessed set of edges induces
a regular subgraph of ${\cal G}$ of degree 3. It is clear that
this whole process can be implemented by a program scheme of
$\mbox{NPSB}(1)$: hence, the result follows.\qed
\end{proof}

The facts that the problem CUB can not be defined in ${\cal
L}^\omega_{\infty\omega}$ (see \cite{SteMLQ}) and that
non-recursive problems can be defined in ${\cal
L}^\omega_{\infty\omega}$ (see \cite{EF95}) immediately yield the
following result.

\begin{corollary}\label{notbddinflogic}
{\em There are problems definable in $\mbox{NPSB}(1)$ \/{\em (}and
so NPSB\/{\em )} which are not definable in ${\cal
L}^\omega_{\infty\omega}$\/{\em ;} and there are problems
definable in ${\cal L}^\omega_{\infty\omega}$ which are not
definable in NPSB.}\qed
\end{corollary}

\section{Ordered structures and amended semantics}

Given our characterization, in the preceding section, of the class
of problems accepted by the program schemes of NPSB, we now
consider the class of problems accepted by these program schemes
when we restrict ourselves to ordered structures.

Using Theorem~\ref{CUB=NP}, we can easily modify the program
scheme implicit in the proof of Lemma~\ref{CUBinNPSB} so that, in
the presence of a built-in successor relation, it accepts any
given problem in {\bf NP}. Conversely, any problem in
$\mbox{NPSB}_s(1)$ is in {\bf NP}. Theorem~\ref{NPSB(1)=Omega_b}
then yields the following result.

\begin{corollary}\label{NP=NPSB_s(1)}
{\em As classes of problems, $\mbox{{\bf NP}}
=\mbox{NPSB\/}_s(1)$\/{\em ;} and $\Omega_b$ is complete for {\bf
NP} via quantifier-free first-order translations with
successor.}\qed
\end{corollary}

By Corollary~\ref{Omegab=NPSB}, $\mbox{NPSB}_s =
(\pm\Omega_b)^\ast[\mbox{FO}_s]$; and by Corollary~5.5 of
\cite{SteHP} and Corollary~\ref{NP=NPSB_s(1)},
$(\pm\Omega_b)^\ast[\mbox{FO}_s] =
(\pm\mbox{HP})^\ast[\mbox{FO}_s]$, where HP is the problem over
the signature $\sigma_{2++}$ consisting of all those
$\sigma_{2++}$-structures ${\cal A}$ which, when considered as
digraphs with edge relation $E^{{\cal A}}$ and two given vertices
$C^{{\cal A}}$ and $D^{{\cal A}}$, are such that there is a
Hamiltonian path from $C^{{\cal A}}$ to $D^{{\cal A}}$.
Furthermore, by Corollary 3.2.2 of \cite{Ste93},
$(\pm\mbox{HP})^\ast[\mbox{FO}_s] = \mbox{{\bf
L}}^{\mbox{\scriptsize{\bf NP}\normalsize}}$ (the class or
problems accepted by a log-space deterministic oracle Turing
machine with access to an {\bf NP} oracle), and every problem in
$(\pm\mbox{HP})^\ast[\mbox{FO}_s]$ can be defined by a sentence
of the form:
$$\exists z_1\exists z_2\ldots \exists z_m(\mbox{HP}[\lambda
{\bf x},{\bf y}\psi({\bf x},{\bf y},{\bf z})]({\bf 0},{\bf max})
\wedge \neg \mbox{HP}[\lambda {\bf x},{\bf y}\varphi({\bf x},{\bf
y},{\bf z})]({\bf 0},{\bf max})),$$ where: ${\bf x}$ and ${\bf
y}$ are $k$-tuples of variables, for some $k$; $\psi$ and
$\varphi$ are quantifier-free first-order formulae (with
successor); and ${\bf 0}$ (resp. ${\bf max}$) is the constant
symbol $0$ (resp. $max$) repeated $k$ times. Hence, translating
this normal form into a program scheme yields that any problem in
$\mbox{NPSB}_s$ can actually be accepted by a program scheme of
$\mbox{NPSB}_s(3)$. Furthermore, any problem accepted by a
program scheme $\forall z_1\forall z_2\ldots\forall z_m\rho$ of
$\mbox{NPSB}_s(2)$ can be accepted by a program scheme of
$\mbox{NPSB}_s(1)$: we simply replace the universal
quantification by code within a program scheme of
$\mbox{NPSB\/}_s(1)$ which uses a while instruction and the
successor relation to check whether a structure is accepted by
$\rho$ for every valuation of the free variables
$z_1,z_2,\ldots,z_m$. Hence, we have the following result.

\begin{theorem}\label{NPSBs}{\em $\mbox{NPSB}_s(1) = \mbox{NPSB}_s(2) =
\mbox{{\bf NP} and } \mbox{NPSB}_s(3) = \mbox{NPSB}_s = \mbox{{\bf
L}}^{\mbox{\scriptsize{\bf NP}\normalsize}}$.\/}\qed
\end{theorem}

Let us now amend our semantics of the program schemes of NPSB.
When we defined the semantics of a program scheme $\rho$ of
$\mbox{NSPB}(2i+1)$, for some $i > 0$, we insisted that when a
test in some if-then-else or while instruction is evaluated
(recall, such a test is a program scheme of $\mbox{NSPB}(2i)$),
the only values used in this evaluation are the current values of
the variables of $\rho$. In particular, all arrays involved in the
evaluation are initialized to $0$ prior to the evaluation.
Suppose that we now insist that arrays used in the evaluation are
{\em initialized to their current values\/} prior to the
evaluation. Consequently, not only can we pass the current values
of the variables across to an evaluation, we can pass the current
values of the arrays across too (or course, when the program
scheme $\rho$ resumes after evaluation of the test, the values of
the arrays are what they were prior to the evaluation of the
test). We denote the program schemes of NPSB with this semantics
as $\mbox{NPSB}^p$ to reflect the fact that a polynomial number of
values is passed across in an evaluation (rather than just a
constant number in the standard semantics). Allowing a polynomial
number of values to be passed across to an evaluation drastically
changes the expressibility of the resulting class of program
schemes (modulo the usual complexity-theoretic qualifications).
The complexity class {\bf PH} is the {\em Polynomial Hierarchy\/};
that is, $\mbox{{\bf PH}} = \cup_{i=1}^\infty\Sigma_i^p$, where
$\Sigma_1^p = \mbox{{\bf NP}}$ and where, for each $i\geq 2$,
$\Sigma^p_i = \mbox{{\bf NP}}^{\Sigma^p_{i-1}}$ (the class of
problems accepted by a polynomial-time non-deterministic oracle
Turing machine with access to a $\Sigma^p_{i-1}$ oracle).

\begin{theorem}{\em $\mbox{NPSB\/}^p(1) = \mbox{NPSB\/}(1)$, $\mbox{NPSB\/}^p(2) =
\mbox{NPSB\/}(2)$ and for every $i\geq 2$, $\mbox{NPSB\/}^p(2i-1)
= \mbox{NPSB\/}^p(2i) = \Sigma^p_i$. Consequently,
$\mbox{NPSB\/}^p = \mbox{{\bf PH}}$.}
\end{theorem}

\begin{proof}
Similarly to the proof (elucidated immediately prior to
Theorem~\ref{NPSBs}) that $\mbox{NPSB}_s(1) = \mbox{NPSB}_s(2)$,
so we can show that $\mbox{NPSB}^p(2i-1) = \mbox{NPSB}^p(2i)$,
for all $i\geq 2$. Obviously, $\mbox{NPSB}^p(1) = \mbox{NPSB}(1)$
and $\mbox{NPSB}^p(2) = \mbox{NPSB}(2)$ (as our original
semantics and our amended semantics do not differ in these cases).

We now show how to build our own successor relation using a
program scheme of $\mbox{NPSB}^p(3)$. Essentially, we guess a
successor relation and store it in the array $S$, of dimension 2,
via the following code:
\begin{tabbing}
\hspace{0.2in}\={\tt $x$ := 0}\\
\>{\tt while $x$ $\neq$ $max$ do}\\
\>\hspace{0.2in}\={\tt guess $y$}\\
\>\>{\tt if $x$ $\neq$ $y$ then}\\
\>\>\hspace{0.2in}\={\tt $S[x,y]$ := $max$}\\
\>\>\>{\tt $x$ := $y$}\\
\>\>\>{\tt fi}\\
\>\>{\tt od}
\end{tabbing}
Then we check, using an if-then-else instruction with the test a
program scheme of $\mbox{NPSB}^p(2)$, that every value appears in
the guessed relation $S$ and that no value appears more than once.
Consequently, by Corollary~\ref{NP=NPSB_s(1)}, any problem in
{\bf NP} can be accepted by some program scheme of
$\mbox{NPSB}^p(3)$.

Not withstanding the above remark, we would like to explicitly
simulate a non-deterministic polynomial-time Turing machine
computation using a program scheme of $\mbox{NPSB}^p(3)$. We can
use arrays to store the work-tape of any such Turing machine and
our successor relation, held in $S$, to mirror the movement of the
tape heads. Our only restriction to this simulation is that we
can only set array values at $max$: we can not reset them to $0$.
Hence, the obvious means of simulation is doomed to failure given
that, in general, the contents of a cell of a Turing machine
work-tape fluctuate and that if we simulate a cell of the work
tape using a fixed number of array elements then we can only
register a constant number of changes to the cell contents.
However, we can get round this difficulty by using the fact that
any (accepting) computation of our Turing machine has length
polynomial in the size of the input structure: hence, we can use
an array to store the complete history of changes to the contents
of a Turing machine work-tape cell as follows.

For simplicity, assume that we wish to hold the contents of $n$
Turing machine work-tape cells (where the input structure has
size $n$) using some arrays and that these contents are only ever
$0$ or $1$. Furthermore, assume that the time taken by our Turing
machine to accept (if it does) is $n$. The general case where a
cell can contain more symbols, where there is a polynomial number
of work-tape cells to deal with and where the Turing machine
accepts in a polynomial number of steps can be handled similarly
by increasing the dimensions of our arrays. Let $A$ and $B$ be
array symbols of dimension 2. Using our successor relation
(constructed earlier), we use the array cells $A[u,1], A[u,2],
\ldots, A[u,n]$ (we think of the elements of our input structure
as being named $\{1,2,\ldots,n\}$ with the names reflecting our
successor relation) to register the first change of the contents
of the work-tape cell $u$, the second change of the work-tape
cell $u$, the third change of the work-tape cell $u$, and so on;
and the array cells $B[u,1], B[u,2], \ldots, B[u,n]$ to register
the value of work-tape cell $u$ after the first change, the value
of work-tape cell $u$ after the second change, the value of
work-tape cell $u$ after the third change, and so on.

If $A[u,i] = max$ then this is interpreted as meaning that there
have been at least $i$ changes of contents; and if $B[u,i] = 0$
(resp. $B[u,i] = max$) then this is interpreted as meaning that
after the $i$th change, the contents of work-tape cell $u$ is $0$
(resp. $1$). Note that when the work-tape cell $u$ changes from
$1$ to $0$, on the $i$th change, say, in order to register this
change we need only set $A[u,i] = max$ and leave $B[u,i]$ alone
(as it has been initialized to $0$). Furthermore, with this
representation, and using our successor relation, we can easily
determine the current contents of any work-tape cell: we simply
cycle down the array $A$ to find the last change of contents and
then ascertain the current contents using $B$. Thus, it should be
clear how we can explicitly simulate our Turing machine
computation using a program scheme of $\mbox{NPSB}^p(3)$.

Now, consider a polynomial-time non-deterministic oracle Turing
machine $M$ consulting an {\bf NP} oracle. By
Corollary~\ref{NP=NPSB_s(1)}, and using an array to hold the
contents of the oracle tape, we can simulate an oracle call of
$M$ by an if-then-else instruction where the test is a program
scheme of $\mbox{NPSB}^p(2)$ (exactly because we are allowed, in
our modified semantics, to pass the values of arrays over to the
evaluation of a test). Hence, we have essentially proven that any
problem in $\mbox{{\bf NP}}^{\mbox{\scriptsize{\bf
NP}\normalsize}}$ can be accepted by a program scheme of
$\mbox{NPSB}^p(3)$. Conversely, it is straightforward to see that
any problem accepted by a program scheme of $\mbox{NPSB}^p(3)$
can be accepted by a polynomial-time non-deterministic oracle
Turing machine with an oracle in ${\bf NP}$ (the only point worthy
of note in this regard is that we must ensure that the contents of
all arrays in the program scheme are written on the simulating
Turing machine's oracle tape). Hence, $\mbox{NPSB}^p(3) =
\mbox{{\bf NP}}^{\mbox{\scriptsize{\bf NP}\normalsize}}$.

The general result now follows by a simple induction: for
example, any polynomial-time non-deterministic oracle Turing
machine consulting an oracle in $\mbox{{\bf
NP}}^{\mbox{\scriptsize {\bf NP}\normalsize}}$ can be explicitly
simulated; and by above the oracle calls can be simulated by
if-then-else instructions where the tests are program schemes
from $\mbox{NPSB}^p(4)$.\qed\end{proof}

\section{Some relative computational capabilities}

We now turn to the relative computational capabilities of the
classes of program schemes $\mbox{NPSB}(1)$ and $\mbox{NPSA}(1)$
on the class of all finite structures (we have more to say about
comparing the classes NPSB and NPSA in the Conclusion).

The following definitions are essential to what follows. Let
$\sigma$ be some signature and let ${\cal A}$ and ${\cal B}$ be
$\sigma$-structures. If $|{\cal A}|\subseteq|{\cal B}|$ and:
\begin{itemize}
\item for every relation symbol $R$ of $\sigma$,
$R^{{\cal A}}$ is $R^{{\cal B}}$ restricted to $|{\cal A}|$; and
\item for every constant symbol $C$ of $\sigma$, $C^{{\cal A}}=C^{{\cal B}}$,
\end{itemize} then we say that ${\cal A}$ is a {\em sub-structure\/} of ${\cal B}$
and write ${\cal A}\subseteq{\cal B}$. If the problem $\Omega$
over $\sigma$ is such that for all $\sigma$-structures ${\cal A}$
and ${\cal B}$ for which ${\cal A}\subseteq{\cal B}$, it is
necessarily the case that ${\cal A}\in\Omega$ implies ${\cal
B}\in\Omega$, then we say that $\Omega$ is {\em closed under
extensions\/}. Let ${\cal EXT}$ be the class of all problems that
are closed under extensions.

\begin{lemma}\label{NPSA1clsedext} {\em Every problem in $\mbox{NPSA\/}(1)$ is
closed under extensions. }
\end{lemma}

\begin{proof} Let $\Omega$ be a problem over the signature $\sigma$ accepted by
the program scheme $\rho$ of $\mbox{NPSA}(1)$. Let ${\cal A}$ and
${\cal B}$ be $\sigma$-structures such that ${\cal
A}\subseteq{\cal B}$, and suppose that ${\cal A}\models \rho$.
Consider the program scheme $\rho$ on input ${\cal B}$ where $0$
and $max$ are chosen to be distinct elements of $|{\cal A}|$. By
`mirroring' an accepting computation of $\rho$ on input ${\cal
A}$, with the chosen $0$ and $max$, we obtain an accepting
computation of $\rho$ on input ${\cal B}$ (the fact that all
tests in while, if and if-then-else instructions are
quantifier-free first-order enables us to do this). Hence, ${\cal
B}\in\Omega$. \qed\end{proof}

\begin{theorem}\label{maintheorem}{\em
$\mbox{{\bf PSPACE}}\cap{\cal EXT}=\mbox{NPSA\/}(1)$ and
\/$\mbox{{\bf NP}}\cap{\cal EXT}=\mbox{NPSB\/}(1)$. }\end{theorem}

\begin{proof} Let $\Omega$ be some problem in $\mbox{{\bf PSPACE}}\cap{\cal EXT}$.
By \cite{SteActa}, there exists a program scheme
$\rho\in\mbox{NPSA}_s(1)$ accepting $\Omega$. Modify $\rho$ to
obtain the program scheme $\rho^\prime\in\mbox{NPSA}(1)$ as
follows. In $\rho^\prime$, begin by guessing a successor
relation; that is, when ${\cal A}$ is some input structure, guess
elements $u_1,u_2,\ldots,u_m\in|{\cal A}|$ so that
$$M[0]=u_1, M[u_1]=u_2,\ldots,M[u_m]=max,$$ where $M$ is a new one-dimensional
array symbol and where the elements of
$\{0,u_1,u_2,$\linebreak$\ldots,u_m,max\}$ are distinct (this
latter condition can be checked as we guess). Replace any atomic
relation of the form $succ(x,y)$ in $\rho$ with the formula
$y=M[x]$, and replace any instruction of the form {\tt guess $x$}
with the following fragment of code:
\begin{tabbing}
\hspace{0.2in}\={\tt guess $x$}\\
\>{\tt $goodx$ := $0$}\\
\>{\tt $ok$ := $0$}\\
\>{\tt while $ok$ = $0$ do}\\
\>\hspace{0.2in}\={\tt if ($x$ = $goodx$ $\vee$ $goodx$ = $max$) then}\\
\>\>\hspace{0.2in}\={\tt $ok$ := $max$}\\
\>\>{\tt else}\\
\>\>\>{\tt $goodx$ := $M[goodx]$}\\
\>\>\>{\tt fi}\\
\>\>{\tt od}\\
\>{\tt if $x$ $\neq$ $goodx$ then `loop forever' fi}
\end{tabbing} (where $goodx$ and $ok$ are new variables). Note that this fragment
of code essentially limits our guesses to elements appearing in
the domain of our guessed successor relation.  We need to show
that acceptance by the program scheme $\rho^\prime$ is invariant
with respect to $0$ and $max$ and that it accepts the problem
$\Omega$.

Suppose that ${\cal A}\in\Omega$. Then ${\cal A}$ is accepted by
$\rho$ no matter which successor relation is chosen for $succ$ in
$\rho$. Choose distinct $0^\prime$ and $max^\prime$ in $|{\cal
A}|$ and a successor relation $succ^\prime$ on $|{\cal A}|$ (with
minimal and maximal elements the chosen elements $0^\prime$ and
$max^\prime$). In particular, $\rho$ accepts ${\cal A}$ with
these constants and this successor relation. Consider a
computation of $\rho^\prime$ on input ${\cal A}$ where the
guessed successor relation is $succ^\prime$. Then there exists a
computation of $\rho^\prime$ mirroring any accepting computation
of $\rho$ on input ${\cal A}$ with this particular successor
relation. That is, ${\cal A}$ is accepted by $\rho^\prime$ and
acceptance does not depend upon the chosen constants $0$ and
$max$.

Conversely, suppose that there is a guessed successor relation,
call it $succ^\prime$ (whose domain need not be all of $|{\cal
A}|$), with minimal and maximal elements $0^\prime$ and
$max^\prime$, yielding an accepting computation of $\rho^\prime$
on input ${\cal A}$. Let $B\subseteq|{\cal A}|$ be the domain of
this successor relation and let ${\cal B}$ be the restriction of
${\cal A}$ to $B$. Then ${\cal B}$ is accepted by $\rho$ when the
successor relation is taken as $succ^\prime$ (note that the
domain of $succ^\prime$ is the whole of $|{\cal B}|$). Hence,
${\cal B}\in\Omega$. However, $\Omega$ is closed under extensions
and so ${\cal A}\in\Omega$. But we have seen from above that if
${\cal A}\in\Omega$ then ${\cal A}$ is accepted by $\rho^\prime$
and acceptance does not depend upon the chosen constants $0$ and
$max$. Thus, acceptance by $\rho^\prime$ is invariant with
respect to $0$ and $max$; and $\mbox{{\bf PSPACE}}\cap{\cal
EXT}\subseteq\mbox{NPSA}(1)$. The fact that every problem in
$\mbox{NPSA}(1)$ can be solved by a polynomial-space algorithm is
straight-forward; and every problem in $\mbox{NPSA}(1)$ is closed
under extensions by Lemma~\ref{NPSA1clsedext}.

Now consider a problem $\Omega\in\mbox{{\bf NP}}\cap{\cal EXT}$
accepted by the program scheme $\rho\in\mbox{NPSB}_s(1)$. We
proceed as above, and define a program scheme
$\rho^\prime\in\mbox{NPSB}(1)$, except with the following
amendment. In $\mbox{NPSB}(1)$, we are only allowed assignments
to array elements of the form {\tt $M[x_1,x_2,\ldots,x_k]$ :=
$max$} and so we need some way of encoding our guessed successor
relation. We encode our relation as:
$$M[0,u_1] = max, M[u_1,u_2] = max, \ldots, M[u_m,max] = max,$$ where $M$ is a new
array symbol of dimension 2. Of course, we ensure that the
elements of $\{0,u_1,u_2,\ldots,u_m,max\}$ are distinct as we
guess. Note that we need to remember the previously guessed
element, $u_i$, so that we know to set $M[u_i,u_{i+1}]$ equal to
$max$. We also need to modify our code so that an atomic relation
of the form $succ(x,y)$ is replaced by the formula {\tt $M[x,y]$ =
$max$}. Arguing as above yields the result. \qed
\end{proof}

One view of Theorem~\ref{maintheorem} is that it provides
syntactic characterizations (via the the classes of program
schemes $\mbox{NPSA}(1)$ and $\mbox{NPSB}(1)$) of semantically
defined complexity classes (namely, $\mbox{{\bf PSPACE}}\cap{\cal
EXT}$ and $\mbox{{\bf NP}}\cap{\cal EXT}$).

\begin{corollary}\label{ExtNPS}{\em
$\mbox{{\bf NP}}=\mbox{{\bf PSPACE}}$ if, and only if,
$\mbox{NPSA\/}(1)=\mbox{NPSB\/}(1)$. }\end{corollary}

\begin{proof} If $\mbox{{\bf NP}}=\mbox{{\bf PSPACE}}$ then $\mbox{{\bf
NP}}\cap{\cal EXT}=\mbox{{\bf PSPACE}}\cap{\cal EXT}$; and so
$\mbox{NPSA}(1)=\mbox{NPSB}(1)$ by Theorem~\ref{maintheorem}.
Conversely, if $\mbox{NPSA}(1)=\mbox{NPSB}(1)$ then
$\mbox{NPSA}_s(1)=\mbox{NPSB}_s(1)$; and so $\mbox{{\bf
NP}}=\mbox{{\bf PSPACE}}$ by \cite{SteActa} and
Corollary~\ref{NP=NPSB_s(1)}.\qed
\end{proof}

Corollary~\ref{ExtNPS} is somewhat surprising given that every
problem in NPSA (and so NPSB) has a zero-one law. In fact, any
problem in ${\cal EXT}$, apart from the empty problem (over some
signature), has a `1-law' (the meaning of `1-law' should be
obvious); and so any non-trivial problem in $\mbox{NPSA}(1)$ (and
$\mbox{NPSB}(1)$) has a 1-law. Note that any class of problems
each of which has a zero-one law can not contain, for example, the
computationally trivial problem consisting of all those
structures of even size.

Corollary~\ref{ExtNPS} can be extended slightly in that we can
obtain some additional equivalences involving fragments of
certain vectorized Lindstr\"{o}m logics. Referring back to
Theorem~\ref{NPSAmain}, the problem mentioned in that theorem is
actually defined as follows.

\begin{definition}
Let the signature $\sigma_{a}=\langle
T_1,T_3,M,C\rangle$, where $M$ is a unary relation symbol, $T_1$ is a binary
relation symbol, $T_3$ is a relation symbol of arity 4 and $C$ is a constant
symbol. We can envisage a $\sigma_{a}$-structure ${\cal A}$ as a Petri net whose
places are given by
$|{\cal A}|$ and whose transitions are given by $T_1$ and $T_2$ via:
\begin{itemize}
\item there is a transition $(\{x\},\{y\})$ whose input place is
$\{x\}$ and whose output place is $\{y\}$ if, and only if, $T_1(x,y)$ holds; and
\item there is a transition $(\{x_1,x_2\},\{y_1,y_2\})$ whose input places are
$\{x_1,x_2\}$ and whose output places are $\{y_1,y_2\}$ if, and only if,
$T_3(x_1,x_2,y_1,y_2)$ holds, where $x_1\not= x_2$ and $y_1\not= y_2$.
\end{itemize} The relation $M$ can be seen as providing an initial marking (with one
token on place $p$ if, and only if, $M(p)$ holds) and the constant $C$ as providing a
final marking (consisting of one token on the place $C$).

A $\sigma_{a}$-structure ${\cal A}$, i.e., a Petri net, complete
with inital and final markings, where every transition has either
2 input places and 2 output places or 1 input place and 1 output
place, is in the problem $\Omega_a$ if, and only if, there is a
marking covering the final marking that is reachable from the
initial marking, i.e., there is a reachable marking in which
there is at least one token on the place $C$.\qed
\end{definition}

\begin{corollary}\label{maincor} {\em The following are equivalent.
\begin{itemize}
\item[{\em (}a\/{\em )}] $\mbox{{\bf NP}}=\mbox{{\bf PSPACE}}$.
\item[{\em (}b\/{\em )}] $\mbox{NPSB\/}(1)=\mbox{NPSA\/}(1)$.
\item[{\em (}c\/{\em )}] $\Omega_b^1[\mbox{FO\/}]=\Omega_a^1[\mbox{FO\/}]$.
\item[{\em (}d\/{\em )}] The problems $\Omega_b$ and $\Omega_a$ are equivalent
via quantifier-free first-order translations with \/{\em 2}
constants.
\end{itemize} }
\end{corollary}

\begin{proof}
Corollary~\ref{ExtNPS} implies ({\em a\/}) $\Leftrightarrow$
({\em b\/}). Theorems~\ref{NPSAmain} and~\ref{NPSB(1)=Omega_b}
imply ({\em b\/}) $\Leftrightarrow$ ({\em d\/}). It is trivially
the case that ({\em c\/}) $\Rightarrow$ ({\em a\/}) and that ({\em
d\/}) $\Rightarrow$ ({\em c\/}).\qed
\end{proof}

We end by returning to an earlier remark concerning the NPSB
hierarchy on the class of all finite structures. We include the
following result here as we can utilize results of this section,
and this result also applies to the NPSA hierarchy.

\begin{proposition}\label{infhier}{\em On the class of all finite structures,
$$\mbox{NPSB\/}(1)\subset\mbox{NPSB\/}(2)\subset\mbox{NPSB\/}(3).$$}
\end{proposition}

\begin{proof}
By Lemma~\ref{NPSA1clsedext}, every problem in $\mbox{NPSB}(1)$
is closed under extensions; and so, trivially,
$\mbox{NPSB}(1)\subset\mbox{NPSB}(2)$.

Consider the following first-order sentence over the signature
$\sigma_2=\langle E,C\rangle$, where $E$ is a binary relation
symbol and $C$ is a constant symbol:
$$\exists x(E(C,x) \wedge \exists y(E(x,y) \wedge \forall z
(E(x,z) \Rightarrow z = y))).$$ There is clearly a program scheme
of NPSB$(3)$ accepting the problem $\Omega$ defined by this
sentence. For any $k\geq 1$, consider the digraphs, ${\cal A}_k$
and ${\cal B}_k$, depicted in
\refstepcounter{fig}\label{ABgraphs}Fig.~\thefig~(note that
${\cal B}_k$ only differs from ${\cal A}_k$ by having an extra
vertex and edge). No matter what the value of $k$, ${\cal
A}_k\in\Omega$ but ${\cal B}_k\not\in\Omega$. We shall show that
for any program scheme $\rho$ of $\mbox{NPSB}(2)$, there exists
some $k$ such that ${\cal A}_k\models \rho$ implies that ${\cal
B}_k\models \rho$. This will yield our result.

\begin{center}
\includegraphics{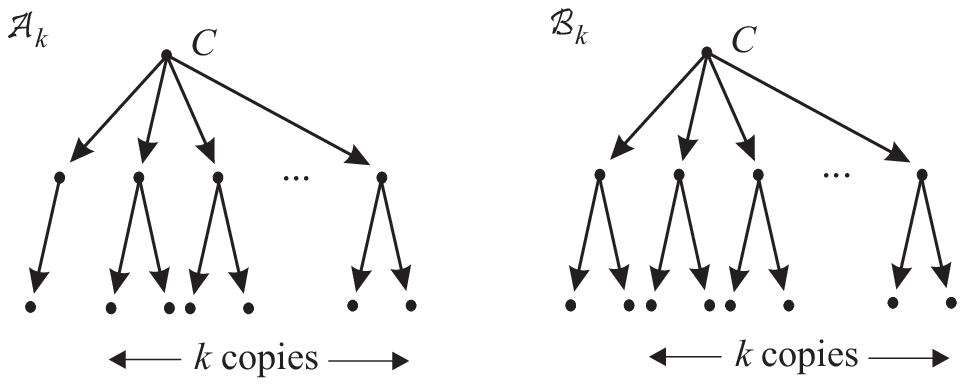}

Figure~\ref{ABgraphs}. The digraphs ${\cal A}_k$ and ${\cal B}_k$.
\end{center}

Let $\rho$ be a program scheme of $\mbox{NPSB}(2)$ of the form
$\forall x_k\forall x_2\ldots\forall x_k\rho^\prime$, for some
program scheme $\rho^\prime$ of $\mbox{NPSB}(1)$, and let
$\tilde{{\cal B}}_k$ be an extension of ${\cal B}_k$ by $k$
constants (one for each variable $x_i$). There is an extension of
${\cal A}_k$, denote it $\tilde{{\cal A}}_k$, such that
$\tilde{{\cal A}}_k$ is embeddable into $\tilde{{\cal B}}_k$ via
a one-to-one mapping: call the mapping $\pi$. Suppose that there
is an accepting computation of $\rho^\prime$ on input
$\tilde{{\cal A}}_k$. We can `mirror' this computation by a
computation of $\rho^\prime$ on $\tilde{{\cal B}}_k$ by making
guesses according to the mapping $\pi$ (after having chosen our
constants $0$ and $max$, again according to $\pi$). The two
computations of $\rho^\prime$, on $\tilde{{\cal A}}_k$ and
$\tilde{{\cal B}}_k$, proceed in tandem (in that their flows of
control are identical) and because the computation of
$\rho^\prime$ on $\tilde{{\cal A}}_k$ leads to acceptance, so
must the computation of $\rho^\prime$ on $\tilde{{\cal B}}_k$
(recall, any tests are quantifier-free first-order and so only
ever refer to the current values of variables). Our result
follows.\qed\end{proof}

We add that the proof of Proposition~\ref{infhier} suffices to
show that, on the class of all finite structures,
$\mbox{NPSA}(1)\subset\mbox{NPSA}(2)\subset\mbox{NPSA}(3).$

\section{Conclusions}

In this paper, we have examined the computational capabilities of
different classes of program schemes, based around `binary
write-once arrays', on the class of finite structures, the class
of ordered finite structures and with respect to different
semantics. We now discuss some potential directions for future
research.

Perhaps the most obvious unanswered question is as regards the
NPSB hierarchy: `Is it the case that, like the NPS and NPSS
hierarchies, the NPSB hierarchy is proper at every level?' (the
same question can be asked for the NPSA hierarchy). So far, we
have not been able to answer this question (beyond
Proposition~\ref{infhier}). The main reason for the lack of
progress is that whereas in \cite{ACS99} we were able to `re-use'
domain elements so as to `mirror' computations of program scheme
of NPS and NPSS (in the style of the proof of
Proposition~\ref{infhier}), the existence of arrays means that we
can `remember the values already used' in a computation and
consequently it is not clear that domain elements can be re-used
in a suitably anonymous fashion (the reader is referred to
\cite{ACS99}, and the proofs therein, in order to make more sense
of this remark). The fact that working with program schemes of
NPSB takes us outside the `bounded-variable world' of the logic
${\cal L}^\omega_{\infty\omega}$ (see
Corollary~\ref{notbddinflogic}), whereas this is not thecase with
the program schemes of NPS and NPSS, is particularly intriguing in
this respect.

The results in Section 6, relating the computational capabilities
of the classes of program schemes $\mbox{NPSB}(1)$ and
$\mbox{NPSA}(1)$, are in the style of Abiteboul and Vianu
\cite{AV89,AV91}, Abiteboul, Vianu and Vardi \cite{AVV97} and
Dawar \cite{Daw98}. However, we would prefer to have determined
similar results but regarding the classes NPSB and NPSA (or,
equivalently, the logics $(\pm\Omega_b)^\ast[\mbox{FO}]$ and
$(\pm\Omega_a)^\ast[\mbox{FO}]$). So far, we have been unable to
extend the results of Section 6 to these classes of programs
schemes. There are some very straightforward implications to be
made however.  For instance (on the class of all finite
structures): \begin{itemize} \item by Corollary~\ref{maincor}, if
${\bf NP} = {\bf PSPACE}$ then $\mbox{NPSB} = \mbox{NPSA}$ (and,
equivalently,
$(\pm\Omega_b)^\ast[\mbox{FO}]=(\pm\Omega_a)^\ast[\mbox{FO}]$);
\item by
Theorems~\ref{NPSAmain} and~\ref{NPSBs}, if $\mbox{NPSB} =
\mbox{NPSA}$ (or, equivalently,
$(\pm\Omega_b)^\ast[\mbox{FO}]=(\pm\Omega_a)^\ast[\mbox{FO}]$)
then $\mbox{{\bf L}}^{\mbox{\scriptsize{\bf NP}\normalsize}} =
{\bf PSPACE}$; and
\item by Theorems~\ref{NPSAmain} and~\ref{NPSBs}, if $\Omega_b^\ast[\mbox{FO}]=\Omega_a^\ast[\mbox{FO}]$
then ${\bf NP} = {\bf PSPACE}$ (as any problem in
$\Omega_b^\ast[\mbox{FO}]$ can easily be seen to be in {\bf NP}).
\end{itemize}
We would like to be able to equate the questions: `Is $\mbox{{\bf
L}}^{\mbox{\scriptsize{\bf NP}\normalsize}}$ equal to {\bf
PSPACE}?', `Is NPSB equal to NPSA?' and `Is
$(\pm\Omega_b)^\ast[\mbox{FO}]$ equal to $
(\pm\Omega_a)^\ast[\mbox{FO}]$?'; as well as the questions: `Is
{\bf NP} equal to {\bf PSPACE}?' and `Is
$\Omega_b^\ast[\mbox{FO}]$ equal to $\Omega_a^\ast[\mbox{FO}]$?'.
As yet, we have been unable to do so.

Finally, let us return to the decision problem described at the
beginning of Section 4 involving the traversal of a digraph
subject to the utilization of user and system resources. We feel
that this problem, and its variations, are very relevant in the
study of the complexity of {\em agent-based systems\/}.
Essentially, an agent-based system is an environment within which
an agent must successfully accomplish a task. Agents interact
with the environment by performing actions and these actions can
result in a change of state of the environment. The reader is
referred to \cite{WD01} for some basic definitions and
complexity-theoretic results in agent-based systems. Our
resource-dependent digraph traversal problem can easily be viewed
as an agent-based system, and we intend to investigate exactly
how the study of program schemes and logics can impact upon that
of agent-based systems in a future paper.

\end{document}